\documentclass[preprint,nofootinbib,superscriptaddress,showpacs]{revtex4-1}

\usepackage{amssymb}
\usepackage{amsmath}
\usepackage{verbatim}
\usepackage{mathrsfs}
\usepackage{amsfonts}
\usepackage{latexsym}
\usepackage{epsfig}
\usepackage{color}
\usepackage{graphicx,subfigure}
\usepackage{units}
\usepackage{slashbox}
\usepackage{natbib}

\begin{document}


\definecolor{orange}{rgb}{0.9,0.45,0} 

\newcommand{\argelia}[1]{\textcolor{pink}{{\bf Argelia: #1}}}
\newcommand{\dario}[1]{\textcolor{red}{{\bf Dario: #1}}}
\newcommand{\juanc}[1]{\textcolor{green}{{\bf JC: #1}}}
\newcommand{\juan}[1]{\textcolor{cyan}{{\bf Juan B: #1}}}
\newcommand{\alberto}[1]{\textcolor{blue}{{\bf Alberto: #1}}}
\newcommand{\miguela}[1]{\textcolor{red}{{\bf Miguel: #1}}}
\newcommand{\miguelm}[1]{\textcolor{orange}{{\bf Megevand: #1}}}
\newcommand{\olivier}[1]{\textcolor{magenta}{{\bf Olivier: #1}}}

\renewcommand{\t}{\times}

\long\def\symbolfootnote[#1]#2{\begingroup%
\def\thefootnote{\fnsymbol{footnote}}\footnote[#1]{#2}\endgroup}

  
\title{Are black holes a serious threat to scalar field dark
  matter models?}

\author{Juan Barranco} \email[]{barranco@astroscu.unam.mx}
\affiliation{Instituto de Astronom\'{\i}a, Universidad Nacional
  Aut\'onoma de M\'exico, Circuito Exterior C.U., A.P. 70-264,
  M\'exico D.F. 04510, M\'exico}

\author{Argelia Bernal} \email[]{argelia.bernal@nucleares.unam.mx}
\affiliation{Instituto de Ciencias Nucleares, Universidad Nacional
  Aut\'onoma de M\'exico, Circuito Exterior C.U., A.P. 70-543,
  M\'exico D.F. 04510, M\'exico}

\author{Juan Carlos Degollado} \email[]{jdaza@astroscu.unam.mx}
\affiliation{Instituto de Astronom\'{\i}a, Universidad Nacional
  Aut\'onoma de M\'exico, Circuito Exterior C.U., A.P. 70-264,
  M\'exico D.F. 04510, M\'exico}

\author{Alberto Diez-Tejedor}
\email[]{alberto.diez@nucleares.unam.mx}
\affiliation{Instituto de Ciencias Nucleares, Universidad Nacional
  Aut\'onoma de M\'exico, Circuito Exterior C.U., A.P. 70-543,
  M\'exico D.F. 04510, M\'exico}

\author{Miguel Megevand}
\email[]{megevand@nucleares.unam.mx}
\affiliation{Instituto de Ciencias Nucleares, Universidad Nacional
  Aut\'onoma de M\'exico, Circuito Exterior C.U., A.P. 70-543,
  M\'exico D.F. 04510, M\'exico}

\author{Miguel Alcubierre}
\email[]{malcubi@nucleares.unam.mx}
\affiliation{Instituto de Ciencias Nucleares, Universidad Nacional
  Aut\'onoma de M\'exico, Circuito Exterior C.U., A.P. 70-543,
  M\'exico D.F. 04510, M\'exico}

\author{Dar\'{\i}o N\'u\~nez}
\email[]{nunez@nucleares.unam.mx}
\affiliation{Instituto de Ciencias Nucleares, Universidad Nacional
  Aut\'onoma de M\'exico, Circuito Exterior C.U., A.P. 70-543,
  M\'exico D.F. 04510, M\'exico}

\author{Olivier Sarbach}
\email[]{sarbach@ifm.umich.mx}
\affiliation{Instituto de F\'{\i}sica y Matem\'aticas, Universidad
Michoacana de San Nicol\'as de Hidalgo, Edificio C-3, Ciudad
Universitaria, 58040 Morelia, Michoac\'an, M\'exico}

\collaboration{Part of the ``Instituto Avanzado de Cosmolog\'ia'' collaboration}


\date{\today}


\begin{abstract} 
Classical scalar fields have been proposed as possible candidates for
the dark matter component of the universe.  Given the fact that
super-massive black holes seem to exist at the center of most
galaxies, in order to be a viable candidate for the dark matter halo a
scalar field configuration should be stable in the presence of a
central black hole, or at least be able to survive for cosmological
time-scales. In the present work we consider a scalar field as a test
field on a Schwarzschild background, and study under which conditions
one can obtain long-lived configurations.  We present a detailed study
of the Klein-Gordon equation in the Schwarzschild spacetime, both from
an analytical and numerical point of view, and show that indeed there
exist quasi-stationary solutions that can remain surrounding a black
hole for {\em large}\/ time-scales.
\end{abstract}




\maketitle


\section{Introduction}
\label{sec:introduction}

Scalar fields are a common theme in modern cosmology. They play a
central role in inflation~\cite{Lidsey:1995np}, and they have been
frequently used to describe dark energy in place of the cosmological
constant~\cite{Peebles:2002gy, Padmanabhan:2002ji}.  Moreover, even if
the usual description of dark matter is done in terms of weakly
interacting massive particles (WIMPs), given the persistent
uncertainty about the real nature of that mysterious component, one
can not {\em a priori}\/ reject one candidate in favor of another.  In
this context, ultra-light scalar fields have also been invoked as
possible alternative dark matter candidates~\cite{Turner:1983he,
  Sin:1992bg, Peebles:1999fz, Peebles:2000yy, Sahni:1999qe, Hu:2000ke,
  Matos:2000ng, Matos:2000ss, Nucamendi:2000jw, Arbey:2001qi,
  Matos:2003pe, Lee:2008jp, Marsh:2010wq, Lundgren:2010sp, Su:2010bj,
  AmaroSeoane:2010qx, Briscese:2011ka, Harko:2011xw}.

In modern physics quantum fields are considered to be the fundamental
constituents of nature. For the particular case of a boson field which
has not been in thermal equilibrium with the other fields of the
standard model of particle physics, a classical field description
(i.e. a coherent excitation) seems in principle as natural as that
given by the WIMP hypothesis.  Indeed, a description in terms of
classical scalar fields has been shown to describe the large scale
cosmological scenario (to linear order) as consistently as cold dark
matter (CDM)~\cite{Hwang:1996xd,Hu:1998kj,Matos:2000ss, Marsh:2010wq}.
In a more local context, massive complex scalar fields can form
spherically symmetric, asymptotically flat, regular, self-gravitating
configurations known as ``boson stars''~\cite{Kaup68,Ruffini69} (see
also \cite{Jetzer92,Schunck:2003kk} and references therein).  For the
case of a scalar field with a mass of the order of $10^{-22}$ -
$10^{-24}$eV, these configurations can have the mass and the size
expected for a typical galactic halo~\cite{Sin:1992bg, Lee:1995af,
  Arbey:2001qi}.  However, it has been recognized that even though
such configurations are stable~\cite{1989NuPhB.315..477L,Seidel90},
they do not reproduce the observed velocity profiles
(though this problem can be significantly alleviated with the use of
multi-state configurations, see e.g.~\cite{Matos:2007zza,
  Bernal:2009zy, UrenaLopez:2010ur}).  Finally, ultra-light scalar
field dark matter (SFDM) could have some advantages over the standard
Lambda cold dark matter ($\Lambda$CDM) scenario, namely that the SFDM
model is not expected to produce an over-density of satellite
galaxies~\cite{Matos:2000ss}, nor does it result in the generation of
the cuspy halos typically obtained in the high-resolution N-body
simulations performed within the standard $\Lambda$CDM
model~\cite{Matos:2003pe,Lee:2008jp,Su:2010bj,Harko:2011xw}.

Despite all this, there has been some controversy regarding the
possibility that dark matter halos could be described by coherent
scalar excitations. The central issue being the fact that most
galaxies seem to host a super-massive black hole at their centers, so
that a scalar field halo should be able to survive surrounding such a
black hole for cosmological time-scales.  Regarding that question, it
has been known for a long time that static, spherically symmetric,
asymptotically flat scalar field configurations with a Schwarzschild
black hole at their center can not exist in nature~\cite{Pena:1997cy},
i.e. a Schwarzschild black hole can not have ``scalar
hair''~\cite{Bekenstein:1995un}. \footnote{Here we will be working
  within the domain of general relativity.  However, ``hairy'' black
  hole configurations are indeed possible in the context of
  scalar-tensor theories of gravity~\cite{Nucamendi:1995ex}.} However,
these ``no-hair'' theorems have nothing to say about a specific {\em
  time-scale}, and it could in principle be possible that scalar field
configurations could survive the birth of a super-massive black hole
and remain in a quasi-stationary state for very long times.
 
The question of how a scalar field is accreted into a Schwarzschild
background has attracted the attention of many researchers, motivated
in some cases by the fact that dark energy can be modeled by a scalar
field~\cite{Marsa:1996fa,Jacobson:1999vr,Bean:2002kx,Frolov:2004vm,
  Babichev:2004yx,Babichev:2005py, UrenaLopez:2011fd}, or because of
the scalar dark matter
models~\cite{UrenaLopez:2002du,CruzOsorio:2010qs}.  A definite picture
of the accretion of a scalar field onto a black hole is not clear yet
and there is in fact some controversy around it.  For instance,
in~\cite{UrenaLopez:2002du} it is argued that the accretion rate will
be very slow, while others argue that this accretion will be
fast~\cite{CruzOsorio:2010qs}. Very long lasting configurations have
also been found using ``artificial'' scalar field
potentials~\cite{Megevand:2007uy}.

Here we will restrict ourselves to the case of a scalar field as a
{\em test field}\/ in the background spacetime of a Schwarzschild
black hole.  It is for this reason that we will talk about ``scalar
field clouds'' rather than ``galactic halos'', in order to maintain a
clear perspective of the real problem we are studying.  The dark
matter content in the central region of a galactic halo is very
diluted, so it is to be expected that this approximation could shed
some light on the nature of the dark matter problem.  However, we
would like to emphasize that this is only a first step on the study of
long-lived scalar field configurations surrounding a black hole, and
the self-gravitating case will be considered in future works.

This paper can be separated in two distinct parts plus the
conclusions. In a first part (Section~\ref{sec:analytical}), we
introduce the problem and present some important analytical results.
We find that it is indeed not possible to obtain a stationary,
localized, {\em physical}\/ scalar field configuration surrounding a
Schwarzschild black hole: Solutions that are well behaved at spatial
infinity are not regular at the horizon; they oscillate with a
divergent frequency and these oscillations produce an infinite
integrated energy close to the horizon.  Thus, in particular, such
solutions are not compatible with the test field limit
approximation. This is of course not surprising if we remember the
no-hair theorems.  However, these {\em non-physical}\/ solutions can
be quite useful in order to construct appropriate initial data for a
special set of {\em physical}\/ quasi-stationary (resonant) scalar
field configurations with very long lifetimes.  In a second part
(Section~\ref{sec:numerical}), we perform numerical evolutions in
order to determine the lifetime of such quasi-stationary resonant
configurations. Finally, we discuss our findings in
Section~\ref{sec:conclusions}.

Throughout the paper we will use geometric units such that $G=c=1$,
and we will use the conventions of Misner, Thorne and Wheeler
(MTW)~\cite{Misner73}.  In particular, the signature of the spacetime
metric will be taken to be ($-,+,+,+$).


\section{Eigenmode analysis and resonances}
\label{sec:analytical}

We will be looking for stationary scalar field configurations
surrounding a Schwarzschild black hole. Here we will restrict
ourselves to the case of a canonical, massive, non-self-interacting,
minimally coupled scalar field $\phi$. The equation of motion for that
field is given by the Klein-Gordon equation
\begin{equation}
(\Box - \mu^2)\phi = 0 \; ,
\label{eq.motion}
\end{equation}
with the d'Alambertian operator defined as $\Box :=
(1/\sqrt{-g})\:\partial_\mu(\sqrt{-g}g^{\mu\nu}\partial_{\nu})$.  With
our conventions $\phi$ is dimensionless, while $\mu$ has dimensions of
(length)$^{-1}$.  The associated quantum mechanical ``mass'' of the
scalar field is given by $\hbar\mu$, but all our considerations in
this work will be purely classical.  (For the SFDM models usually
considered in the literature one usually has \mbox{$\hbar\mu\sim$
  $10^{-22}$ - $10^{-24}$eV} in physical units, which implies $\mu$ of
the order of \mbox{$1$ - $10^{-2}$(pc)$^{-1}$}.)

We will also be assuming that the energy associated with the scalar
field configuration is very small compared to the mass of the black
hole, so that the gravitational back-reaction associated to the scalar
field distribution can be disregarded.  Under this assumption, and
working in Schwarzschild coordinates, we can write the spacetime
metric as
\begin{equation}
ds^2 = - N(r) dt^2 + \frac{dr^2}{N(r)} + r^2 d\Omega^2 \; , \qquad
N(r) := 1 - 2M/r \; ,
\label{Schwarzschild}
\end{equation}
with $M$ the mass of the black hole and $d\Omega^2 := d\theta^2 +
\sin^2\theta d\varphi^2$ the standard solid angle element.  This
coordinate system is suitable to describe the physics in the exterior
region $r\in(2M,\infty)$.

To look for solutions of the Klein-Gordon equation above, we will
start by considering a decomposition into spherical
harmonics:~\footnote{If the scalar field was real we would have
  $\psi_{\ell, -m}=(-1)^{m}\psi^{*}_{\ell, m}$, i.e. not all the
  modes $\psi_{\ell m}(t,r)$ in~\eqref{decomposition} are independent
  of each other.  Apart from this, all the results obtained below
  apply for the real as well as the complex cases.  The differences
  between the real and complex scalar field will appear when
  considering the energy-momentum tensor, but this is not relevant for
  the test field limit approximation we are considering here.}
\begin{equation}
\phi(t,r,\theta,\varphi)  = \frac{1}{r}\sum\limits_{\ell, m} \psi_{\ell m}(t,r)
Y^{\ell m}(\theta,\varphi) \; ,
\label{decomposition}
\end{equation}
where the $1/r$ factor has been introduced for convenience, and the
parameters $\ell$ and $m$ take the usual values: $\ell \in
\{0,1,2,\ldots \}$, $-\ell\le m \le \ell$.  Substituting the above
equation into~\eqref{eq.motion} we obtain the following family of
reduced equations
\begin{equation}
\left[ \frac{1}{N(r)}\frac{\partial^2}{\partial t^2} 
 - \frac{\partial}{\partial r} N(r)\frac{\partial}{\partial r}
 + \mathcal{U}_{\ell}(\mu, M; r) \right]\psi_{\ell m} = 0 \; ,
\label{Eq:ReducedKG}
\end{equation}
where we have defined
\begin{equation}\label{effective.potential}
\mathcal{U}_{\ell}(\mu, M; r) := \frac{\ell(\ell+1)}{r^2}
+ \frac{2M}{r^3} + \mu^2 \; .
\end{equation}
Notice that $\mathcal{U}_{\ell}$ is an everywhere positive and
monotonically decreasing function of $r$, and also that
equation~\eqref{Eq:ReducedKG} does not depend on $m$.  The first term
in~\eqref{effective.potential} is the usual centrifugal term, the
second one is a curvature correction, and the third one is inherited
from the corresponding term in the Klein-Gordon equation,
Eq.~\eqref{eq.motion}.

In the test field limit all configurations have the conserved energy
\mbox{$E=\sum\limits_{\ell, m} E_{\ell m}$}, with
\begin{subequations}\label{energy}
\begin{equation}
E_{\ell m} = \int_{2M}^\infty \rho_E(r) dr\; , 
\end{equation}
and where
\begin{equation}
\rho_E (r)=  \frac{1}{2} \left( 
 \frac{1}{N(r)}\left| \frac{\partial\psi_{\ell m}}{\partial t} \right|^2
 + N(r)\left| \frac{\partial\psi_{\ell m}}{\partial r} \right|^2
 + \mathcal{U}_{\ell}(\mu, M; r) |\psi_{\ell m}|^2 \right)
\end{equation}
\end{subequations}
is the energy density integrated over the sphere as a function of
areal radius.  In what follows we will refer to this integrated
energy density simply as the ``radial energy density''.

To obtain the last expression we have used the fact that it is always
possible to associate a conserved charge $Q_a=\int_\Sigma
k_{a}{}^{\mu}T_{\mu\nu}n^{\mu}\sqrt{\gamma} \: d\Sigma$ to every
Killing vector field $k_a$.  Here $n^\mu$ is a time-like,
future-directed, normalized 4-vector orthogonal to the 3-dimensional
space-like hypersurface $\Sigma$, and $\sqrt{\gamma} \: d\Sigma$ is
the volume element on $\Sigma$. The subindex $a$ labels the different
Killing fields associated with the background metric, and $T_{\mu\nu}$
is the energy-momentum tensor for the matter fields. Here we will be
only interested in those configurations with a finite value for the
integrals above. Under this assumption and the additional requirement
of $\Sigma$ being a Cauchy surface, it follows that the charges $Q_a$
do not depend on the particular choice of $\Sigma$; in particular the
charges are preserved under time evolution. For the case $k_0 =
\partial_t$ and $\Sigma = \{ t = \textrm{const.} \}$, where
$(t,r,\theta,\varphi)$ are the usual Schwarzschild coordinates, we
arrive at expressions~\eqref{energy}. The hypersurfaces $\Sigma$ are
Cauchy surfaces for the Schwarzschild exterior spacetime, and as a
consequence the energy~\eqref{energy} is conserved. However, in the
numerical simulations of Section~\ref{sec:numerical} below, we will be
integrating the energy density over a different foliation which
extends from the future horizon to some finite, large radius, and the
resulting energy will, in fact, decay due to energy loses through the
horizon and the outer boundary.


\subsection{Reduction to a time-independent Schr\"odinger-like problem:
mode solutions}
\label{sec:schoedinger}

In order to look for the stationary solutions of
Eq.~\eqref{Eq:ReducedKG} we make a further decomposition of the
functions $\psi_{\ell m}(t,r)$ into oscillating modes of the form:
\begin{equation}
\psi_{\ell m}(t,r) = e^{i\omega_{\ell m} t} u_{\ell m}(r) \; ,
\label{Eq:ModeAnsatz}
\end{equation}
with $\omega_{\ell m}$ a real frequency, and $u_{\ell m}(r)$ a complex
function of $r$ in the interval $(2M,\infty)$. In order to simplify
the notation from now on we will omit the labels $\ell$ and $m$ from
$\omega_{\ell m}$ and $u_{\ell m}$.

Introducing the ansatz~\eqref{Eq:ModeAnsatz} into
equation~\eqref{Eq:ReducedKG} one finds the following eigenvalue
problem
\begin{equation}
\left[ - N(r) \frac{\partial}{\partial r} \left( N(r)
\frac{\partial}{\partial r} \right)
+ N(r) \: \mathcal{U}_{\ell}(\mu, M; r) \right] u(r)
= \omega^2 u(r) \; ,\qquad
2M < r < \infty \; .
\label{Eq:EigenvalueProblem}
\end{equation}
After defining the Regge-Wheeler tortoise coordinate \mbox{$r^{*}:=
  r + 2M\ln(r/2M-1)$}, this equation can be rewritten as the
following time-independent Schr\"odinger-like equation:
\begin{equation}
\left[ -\frac{\partial^2}{\partial r^{*2}} 
+ V_{\textrm{eff}}(r^*) \right] u(r^*) = \omega^2 u(r^*) \; ,\qquad
-\infty < r^* < \infty,
\label{Eq:TimeIndependentSchrodinger}
\end{equation}
with the ``effective potential'' $V_{\textrm{eff}}(r^*)$ defined as
\begin{equation}
V_{\textrm{eff}}(r^*) := N(r) \: \mathcal{U}_{\ell}(\mu,M; r) \; , \qquad r=r(r^*).
\label{effectivepotential}
\end{equation}
Notice that in the usual non-relativistic quantum mechanical case one
would have $\omega$ instead of $\omega^2$ in
Eq.~\eqref{Eq:TimeIndependentSchrodinger}.  This is because the
Schr\"odinger equation is first order in time, whereas the
Klein-Gordon equation is second order. Also, note that
equation~\eqref{Eq:TimeIndependentSchrodinger} in fact only depends on
the parameters $\ell$, $M\mu$ and $M\omega$. Since that equation 
is invariant under $\omega \to -\omega$ we can restrict ourselves to the case $\omega>0$.

\begin{figure}
\begin{center}
\includegraphics[width=0.8\textwidth,clip]{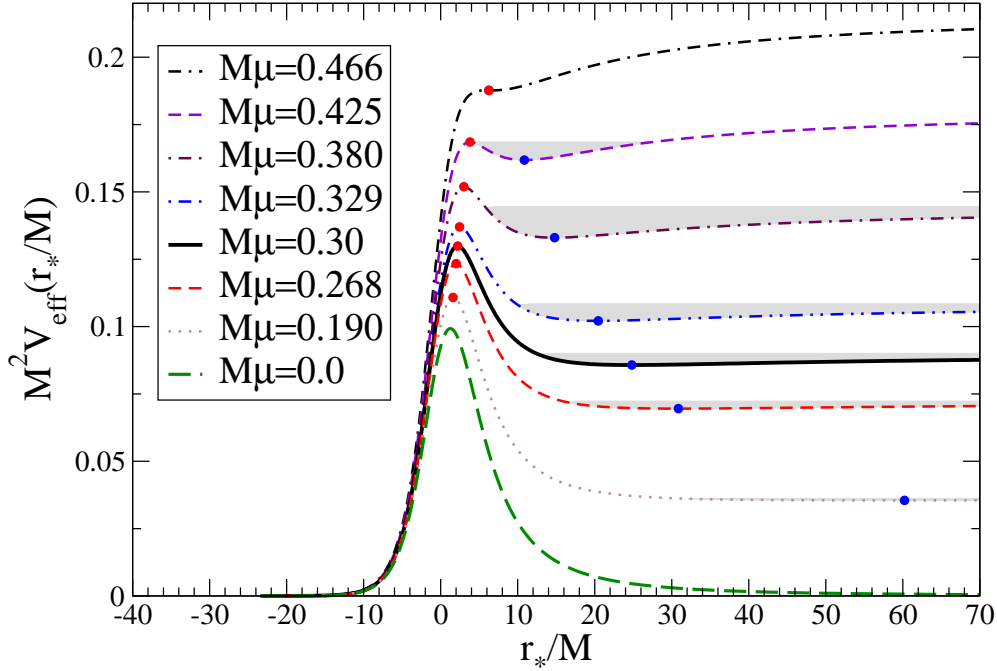}
\caption{The effective potential $M^2 V_{\textrm{eff}}$ for $\ell=1$,
  and different values of the parameter $M\mu$. The shaded regions
  represent the ``resonance bands'' where the potential has a local
  minimum (see Section~\ref{sec:resonant} below).  Notice that for
  $M\mu=0$ and $M\mu \gtrsim 0.466$ the local minimum does not
  appear. For the case $M\mu=0$ we recover the usual Regge-Wheeler
  potential. The case with $M\mu=0.3$ will be considered later on in
  our future examples and it appears as a solid line to emphasize it.}
\label{fig.potentials}
\end{center}
\end{figure}

The typical form of the effective potential can be seen in
Fig.~\ref{fig.potentials}.  Contrary to what happens in the usual
Regge-Wheeler equation, the asymptotic value of $V_{\textrm{eff}}$ for
large radii is not zero but $\mu^2$. This fact will turn out to be
crucial in the study of the behavior of the solutions to
equation~\eqref{Eq:TimeIndependentSchrodinger}.

In the following, we will analyze the spectrum of the Schr\"odinger
 operator given above in more detail.  The operator can be separated
into the point and the continuous spectrum. The point spectrum
consists of the eigenvalues of
Eq.~\eqref{Eq:TimeIndependentSchrodinger} for which the square norm
\begin{equation}
\left\| u \right\|^2:=\int_{2M}^\infty \left( 
 N(r)\left| \frac{\partial u}{\partial r} \right|^2
+ \mathcal{U}_{\ell}(\mu, M; r) |u|^2 \right) dr 
\label{norm}
\end{equation}
is finite. Such solutions, when they exist, describe bound states in
the quantum mechanical picture. In the context of the Klein-Gordon
equation they will give rise to stationary modes \mbox{$\psi(t,r) =
  e^{i\omega t} u(r)$} with finite energy. However, as we will show
below, such solutions do not exist in our case.

This leaves us with the continuous spectrum consisting on the values
of $\omega^2$ for which the time-independent Schr\"odinger
equation~\eqref{Eq:TimeIndependentSchrodinger} admits solutions which
do not grow faster than a polynomial as $r^*\to\pm\infty$. These
solutions describe scattering states and will be analyzed in more
detail below.

At this point it is important to mention that exact solutions to
equation~\eqref{Eq:TimeIndependentSchrodinger} are in fact known and
can be expressed in terms of the so-called ``Heun functions'' (see
e.g.~\cite{Jensen:1985in,Zecca:2009zz}).  However, we will not
use such exact solutions here and will rather concentrate on the
general behavior of the solutions.

We will start with an analysis of the asymptotic behavior of the
solutions to Eq.~\eqref{Eq:TimeIndependentSchrodinger}, both close to
the horizon and close to infinity.


\subsection{Asymptotic behavior and spectrum}
\label{sec:asymptotics}

Regarding the behavior of the effective potential notice first that,
as already mentioned above, in the limit $r^*\to\infty$ it approaches
the constant value $\mu^2$, so that
Eq.~\eqref{Eq:TimeIndependentSchrodinger} reduces to
\begin{equation}\label{eq.r.infty}
\frac{\partial^2 u}{\partial r^{*2}} + k^2 u = 0 \; ,\qquad
k^2 := \omega^2 - \mu^2 \; .
\end{equation}
The general solution to this equation is a linear combination of the
functions $u=e^{\pm ik r^*}$, with $k := \sqrt{\omega^2 -
  \mu^2}$. Therefore, as long as $0<\omega^2 < \mu^2$ the parameter
$k$ will be purely imaginary and we will have one exponentially
decaying and one exponentially growing solution, while for $\omega^2 >
\mu^2$ the parameter $k$ will be real and both solutions will be
oscillatory.

For $r^*\to -\infty$, on the other hand, the effective potential
$V_{\textrm{eff}}$ decays exponentially fast to zero since $N(r) = 2M
r^{-1}( r/2M - 1) = 2M r^{-1}\exp( (r^* - r)/2M )$. Therefore, the
solutions are oscillatory for all $\omega^2 > 0$ (notice that we are
assuming $\omega$ to be real so this is always the case), and they can
be expressed as a linear combination of the functions $u= e^{\pm
  i\omega r^*}$.

Consider now the interval $0 < \omega^2 < \mu^2$. For $\omega^2$ to
belong to the spectrum, the right asymptotic solution must be
proportional to $e^{-|k| r^*}$, and the left asymptotic solution may be
a linear combination of $e^{\pm i\omega r^*}$ of the form
\begin{equation}
u \propto e^{-i\omega r^*} + R e^{i\omega r^*} \; ,
\label{Eq:Reflection}
\end{equation}
with $R$ a reflection coefficient. However, since the right asymptotic
solution and the coefficients in
Eq.~(\ref{Eq:TimeIndependentSchrodinger}) are real, so must be the
left asymptotic solution, which implies that $|R|=1$ such that
$u \propto \cos(\omega r^* - \delta)$ for some phase factor $\delta$. In
particular, the solution is not normalizable with respect to the
square norm given in (\ref{norm}), and so it does not give rise to a
bound state. Therefore, the spectrum for $0 < \omega^2 < \mu^2$ is
continuous and non-degenerate. From the point of view of scattering
theory, $|R|=1$ simply means that a plane wave which travels from
$r^*\to -\infty$ to the right with frequency $0 < \omega^2 < \mu^2$ is
completely reflected from the potential since it does not have enough
``energy" to reach $r^*\to +\infty$. For the Klein-Gordon equation,
this means that the corresponding stationary mode solution satisfies
$\psi(t,r) \propto e^{i\omega(t - r^*)} + R e^{i\omega(t + r^*)}$ for
$r^*\to -\infty$. Since $R\neq 0$, the solution is always singular at
both the future and past horizons.

Finally, for $\omega^2 > \mu^2$ the asymptotic solutions at
$r^*\to\pm\infty$ are both oscillatory, and the spectrum is continuous
and degenerate since there are two linearly independent solutions
which are bounded at $r^*\to\pm\infty$.

Notice that the mode solutions with $0 < \omega^2 < \mu^2$ do not have
a finite value for the energy, expression~\eqref{energy}, but they are
still ``localized'' around the black hole since they decay
exponentially at large $r^*$.  This is not the case for the mode
solutions with $\omega^2 > \mu^2$, which do not decay at spatial
infinity.


\subsection{Resonant States}
\label{sec:resonant}

We will now take a closer look at the energy interval $0 < \omega^2 <
\mu^2$. Even though the reflection coefficient is one in this case
(see Eq.~\eqref{Eq:Reflection} above), it is still possible to have
resonances if $V_{\textrm{eff}}(r^*)$ has a ``potential well'' with a
local minimum lying below $\mu^2$. In this case, plane waves emanating
from $r^*\to -\infty$ may tunnel through the potential barrier into
the potential well, and if they have the correct frequency their
amplitude may build up through multiple reflections at the walls of
the potential well. As we will show below, such resonances indeed
occur for a discrete set of energy levels.

Clearly, a necessary condition for this phenomena to take place is the
existence of the potential well. In order to find under which
conditions such a potential well exists we first need to determine the
critical points of the effective potential $V_{\textrm{eff}}$.  It is
in fact not difficult to show that such critical points correspond to
the roots of the following third degree polynomial in $r$ (notice that
here we are working with $r$ and not with $r^*$):
\begin{equation}
M \mu^2 r^3 - \ell \left( \ell + 1 \right) r^2
+ 3M \left( \ell^2 + \ell - 1 \right) r + 8 M^2 = 0 \; .
\label{thirdorder}
\end{equation}
This polynomial can be shown to have three real solutions if the
following condition is verified:
\begin{eqnarray}
(M \mu)^2 &<& - \frac{1}{32}(\ell^2+\ell - 1)(\ell^2+\ell + 1)^2 \nonumber  \\
&+& \frac{1}{288}\sqrt{3(3\ell^4+6\ell^3+5\ell^2+2\ell+3)^3} \; ,
\label{condition_real}
\end{eqnarray}

If $M\mu$ satisfies relation~\eqref{condition_real} above, the
potential $V_{\textrm{eff}}$ will have three critical points.  One of
them always corresponds to a negative value of $r$, so it is not of
interest to our problem (negative values of $r$ are unphysical).  The
other two critical points correspond to a minimum and a maximum of the
effective potential.

For the case $\mu=0$ we recover the well-known solutions for which the
maximum of the potential is located at:
\begin{equation}
r_{\textrm{max}} = \frac{3M}{2}\left( 1 + \sqrt{1
+ \frac{14\ell^2+14\ell+9}{9\ell^2(\ell+1)^2}}
- \frac{1}{2\ell(\ell+1)} \right) \; .
\end{equation}
For $\ell \rightarrow \infty$ the maximum occurs at $r_{\textrm{max}}=3M$,
which represents the innermost stable circular orbit for massless
classical particles.  Any particle that starts with vanishing radial
velocity in the region $r < 3M$ will fall into the black hole.

\begin{figure}
\begin{center}
\includegraphics[width=0.8\textwidth,clip]{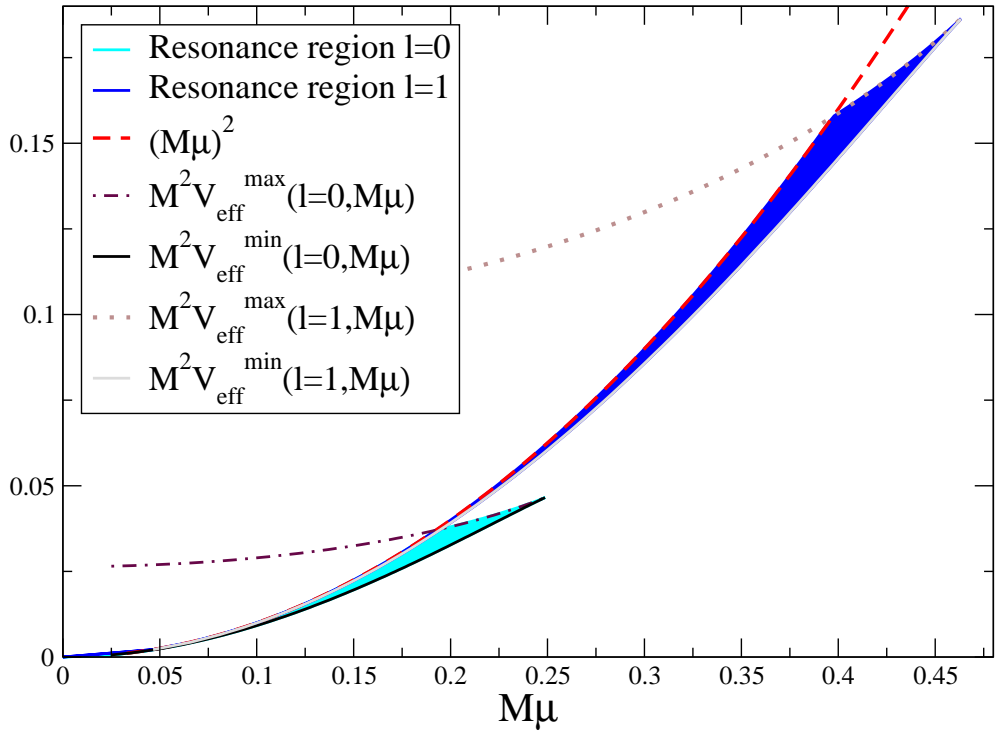}
\caption{Resonance band in parameter space. The cut-off in $M\mu$ is
  reached for values so large that condition~\eqref{condition_real} is
  no longer satisfied.  For $\ell=0$ this condition is $M\mu < 1/4$,
  while for $\ell=1$ one finds \mbox{$(M\mu)^2<-9/32+\sqrt{20577}/288
    \approx 0.217$} , which implies $M \mu \lesssim 0.466$.}
\label{resonance_regions}
\end{center}
\end{figure}

The region in parameter space where a potential well exists and
resonances are possible is illustrated in Fig.~\ref{resonance_regions}
for different values of $\ell$ (see also Fig.~\ref{fig.potentials}).
We will call this region the ``resonance band'', $V_{\rm eff}^{\rm
  min}<\omega^2<\rm{min}\left\{V_{\rm eff}^{\rm max},\mu^2\right\}$,
although not all the states in that band will be resonant.  Here
$V_{\rm eff}^{\rm min}$ and $V_{\rm eff}^{\rm max}$ are the local
minimum and maximum, respectively, represented with dots in
Fig.~\ref{fig.potentials}. The cut-off in $M\mu$ that appears for each
value of $\ell$ corresponds to a value of $M\mu$ large enough so that
relation~\eqref{condition_real} is no longer satisfied and the local
minimum of the potential does not appear. Notice that such a cut-off
increases with $\ell$, so that for a mass of the scalar field such
that $M\mu <1/4$ the effective potential will have a minimum for all
possible values of $\ell$.~\footnote{Taking the mass of a typical
  black hole in the center of a galaxy to be of the order $M\sim
  10^{8}M_{\odot}$ we obtain that the mass of the scalar field should
  be less than $10^{-18}$ eV. Compare that result with the expected
  mass for the axion, $10^{-5}$ to $10^{-3}$ eV \cite{Amsler:2008zzb},
  and with the mass of the SFDM models, $10^{-22}$ to $10^{-24}$eV.}
We also want to stress the fact that we can have a minimum in the
effective potential for $\ell =0$, but not for $\mu=0$. This is the
reason why the phenomenology of the problem we are analyzing here is
quite different to that encountered in the usual Regge-Wheeler
equation.

\vspace{5mm}

Having found under which conditions there is a potential well, we now
turn our attention to the existence of resonant modes. In order to
have modes that are ``almost trapped'' in the potential well it is
clear that one must ask for $0<\omega^2 < \mu^2$. This will guarantee
that the solution for the scalar field decays exponentially at spatial
infinity and is ``localized'' close to the black hole, but of course
the scalar field can still ``escape'' towards the black hole horizon.
The situation we are considering is as follows: We are solving the
stationary Klein-Gordon equation, so close to the horizon (where $r^*$
approaches $-\infty$) we must have a combination of incoming and
outgoing modes of the same amplitude.  This implies that any scalar
field that ``escapes'' from the potential well (via ``quantum
tunneling") is precisely compensated by the incoming field from the
horizon.  The situation is clearly not physical since, first, we do
not expect anything coming from the horizon, and second, the energy of
such solutions is infinite as discussed above, but nevertheless these
are the type of solutions we are considering here.

Now, as already remarked, the frequency spectrum is continuous, so we
will find a solution with the properties just described for any
$0<\omega^2 < \mu^2$.  Clearly, if we want to find solutions that are
in some sense ``trapped'' inside the potential well, we must also ask
for $\omega^2$ to be in the resonance band, \mbox{$V_{\rm eff}^{\rm
    min}<\omega^2<\rm{min}\left\{V_{\rm eff}^{\rm
    max},\mu^2\right\}$}.  The question now is what distinguishes the
set of {\em discrete}\/ resonant solutions from all other solutions in
the continuous spectrum of the resonance band.  There are several ways
in which one can look for such resonant states, but here we have
settled on a practical ``intuitive'' approach: Resonant states will
correspond to those that ``leak out'' of the potential well very
slowly, or in other words, they will correspond to those special
solutions for which the ratio between the amplitude outside the
potential well and the amplitude inside is a minimum.

In order to have a clear picture of what we are talking about here,
let us consider the particular case $\ell=1$, $M\mu=0.3$, which
clearly satisfies the conditions for a potential well to exist.
Figure~\ref{fig.potentials} shows the effective potential $M^2
V_{\textrm{eff}}$ in that case. Notice that there is a clear
minimum at $r^*/M \approx 25$ for which $M^2 V_{\rm eff}^{\rm min} \approx
0.085$, while the maximum is at $r^*/M \approx 2.2$ with $M^2 V_{\rm
  eff}^{\rm max} \approx 0.13$.

We can now solve the Schr\"odinger
equation~\eqref{Eq:TimeIndependentSchrodinger} numerically for an
arbitrary value of $0<\omega^2 < \mu^2$.  The details for the
numerical solver are not particularly important, it is sufficient to
say that the solver starts at a large value of $r^*/M$, for which
$V_{\textrm{eff}} \approx \mu^2$, and where one can approximate the
solution with a decaying exponential.  One then solves the equation
numerically from right to left.

\begin{figure}
\begin{center}
\includegraphics[angle=270,width=0.8\textwidth,clip]{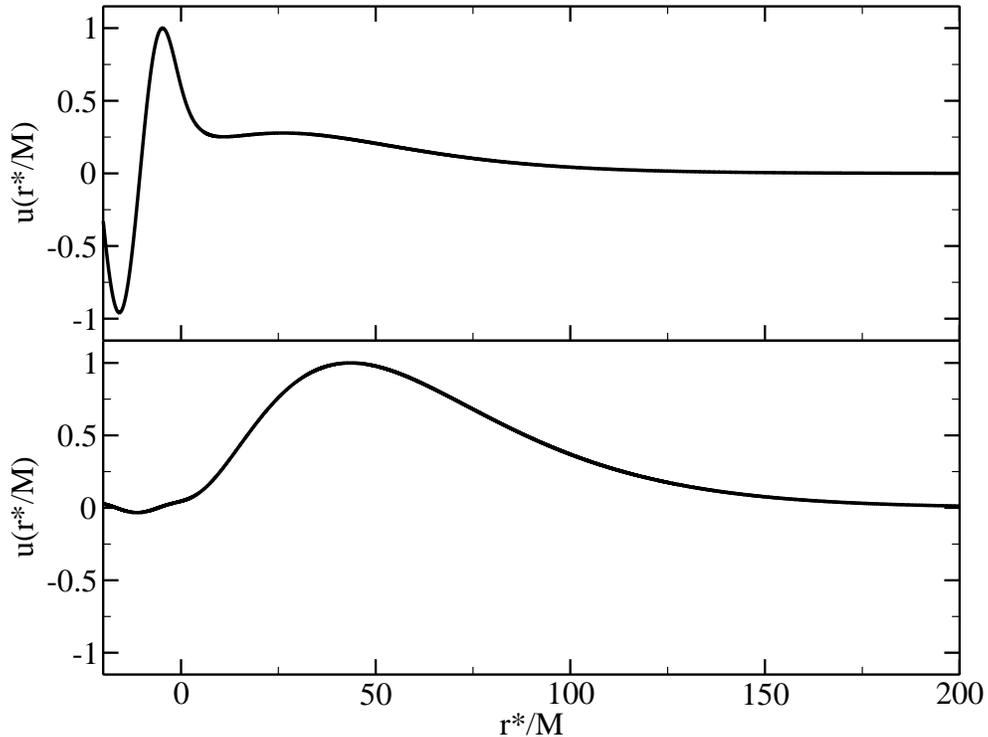}
\caption{Numerical solutions of the Schr\"odinger
  equation~\eqref{Eq:TimeIndependentSchrodinger} for different values
  of $M\omega$.  The upper panel corresponds to $M\omega=0.295$, while
  the lower panel corresponds to $M\omega=0.29619$. Both solutions have
  been scaled so that the maximum overall amplitude is equal to $1$.}
\label{fig:solutions}
\end{center}
\end{figure}

Figure~\ref{fig:solutions} shows two solutions for particular values
of $M\omega$.  The first plot corresponds to the solution for
$M\omega=0.295$.  Notice that for $r^*/M<0$ the solution is
oscillatory as expected, and that it penetrates only slightly into the
potential well: The amplitude for $r^*/M<0$ is much larger than the
amplitude inside the well.  This behavior is quite generic. The
second plot, on the other hand, correspond to the solution for
$M\omega=0.29619$.  Notice that now the situation is quite the
opposite: The amplitude inside the well is much larger than the
amplitude for $r^*/M<0$ (both solutions have been scaled so that the
maximum overall amplitude is equal to $1$).  This means that the value
$M\omega=0.29619$ is close to a resonant frequency; in fact it is
close to the very first resonance, the one with no nodes inside the
potential well.

\begin{figure}
\begin{center}
\includegraphics[angle=270,width=0.8\textwidth,clip]{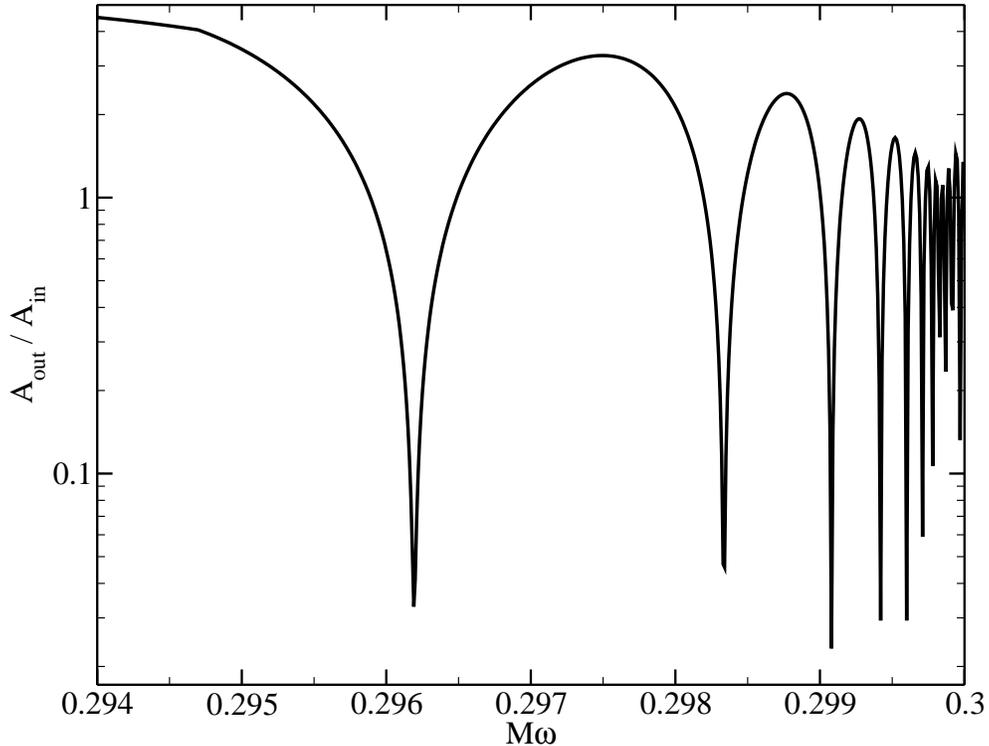}
\caption{Resonant frequencies.  We show a log plot of the ratio
  $A_{\rm out}/A_{\rm in}$ between the amplitude $A_{\rm out}$ of the
  solution for $r^*/M<0$, and the amplitude $A_{\rm in}$ for the
  solution inside the potential well $r^*/M>10$. }
\label{fig:omegamap}
\end{center}
\end{figure}

In order to locate the different resonant frequencies one can find the
solution for all frequencies in the interval $V_{\rm eff}^{\rm min} <
\omega^2 < \mu^2$, and simply plot the ratio between the amplitude
$A_{\rm out}$ of the solution for $r^*/M<0$, and the amplitude $A_{\rm
  in}$ for the solution inside the potential
well. Figure~\ref{fig:omegamap} is just such a plot for the case we
are considering here. Notice that the ratio $A_{\rm out}/A_{\rm in}$
drops to very small values at a set of sharply defined discrete
frequencies.  Each subsequent resonant frequency can be easily shown
(by just plotting them) to correspond to solutions with more nodes
inside the potential well.  As the value of $\omega$ approaches the
parameter $\mu$ the resonant frequencies become closer to each other,
and in fact there seem to be an infinite number of them.  Care must be
taken when finding the numerical solution, since higher resolution is
required as $\omega$ approaches $\mu$ (both in $r^*/M$ and in
$\omega$), while at the same time one needs to consider much larger
values of $r^*/M$ since the solutions become so much wider (specific
examples of such resonant solutions will be considered in the
following sections).

\section{Long Lasting Configurations. Numerical Study}
\label{sec:numerical}

According to the arguments given in Section~\ref{sec:analytical}
above, we can not have stationary, localized, physical scalar field
configurations surrounding a Schwarzschild black hole: The stationary
solutions, even the resonant ones, have a divergent energy due to
their oscillatory behavior close to the horizon.~\footnote{Moreover,
  such solutions are not compatible with the test field limit
  approximation.}  However, one can always construct physical
configurations that are arbitrarily close to the stationary solutions
and that can survive the black hole for arbitrarily long times: Simply
choose any (non-physical) stationary solution that decays at spatial
infinity (i.e.  $0<\omega^2<\mu^2$), and set it to zero by hand in the
interval $r \in \left( 2M,2M+\epsilon \right)$, for some small
parameter $\epsilon>0$ with dimensions of length. 
The resulting configuration can be seen as a combination of the
stationary solution plus a perturbation around the horizon, and it
will be stationary for the domain of dependence of
$[2M+\epsilon,\infty)$.

For a given value of $\epsilon$, and some $\Delta > \epsilon$, it will
take a time of order $t_{\epsilon} (2M+\Delta) = 2M \ln
(\Delta/\epsilon)+(\Delta-\epsilon)$ for the perturbation to reach the
radial coordinate $r=2M+\Delta$.  That is, the standard flat
space-time result $\Delta-\epsilon$, plus an extra term coming from
the non-trivial space-time curvature, $2M \ln (\Delta/\epsilon)$.  In
the limit $\epsilon \to 0$ that time can be made arbitrarily
large. Since the energy density diverges as $r\rightarrow2M$, one
could naively think that we can not make $\epsilon$ arbitrarily small
if we want to remain within the test field approximation, but notice
that, since the Klein-Gordon equation is linear in $\phi$, for any
given value of $\epsilon$ one can always reduce the amplitude of the
scalar field arbitrarily. But of course, that would yield scalar field
distributions in which most of the energy is concentrated very close
to the horizon.

If we take $2M+\Delta$ to be of the order of the size of a given
scalar field distribution surrounding a Schwarzschild black hole, one
would expect that $t_{\epsilon} (2M+\Delta)$ could give us some
insight into the lifetime of that particular
configuration.~\footnote{In fact, for the configurations we will be
  considering here (see Table~\ref{table1}), this lifetime will not be
  particularly sensitive to the parameter $\epsilon$. If the width of
  the scalar field configuration is of order $10^3 M$, then one needs
  to take $\epsilon \lesssim 10^{-11}$ for the gravitational term
  $2M \ln (R/\epsilon)$ to be relevant in the expression for
  $t_\epsilon$.}  It is worth noticing, however, that $t_{\epsilon}$
works only as a lower bound, and some scalar field distributions could
very well last much longer than that. In fact, we will show that
initial data can be constructed in a particular way that produces long
lasting, quasi-stationary scalar field distributions in a compact
region outside the black hole. These initial data correspond to the
resonant states, and we will devote most of the remaining of this
paper to study them, although we will also consider a few other
configurations for comparison.  We will come back to the study of the
characteristic time-scales associated to these configurations in
Section \ref{sec:times}.


\subsection{Initial Data}
\label{sec:initial}

We will now construct pseudo-stationary configurations by truncating
the stationary solutions as explained in the previous section. These
configurations will then be used as initial data to be evolved
numerically so that we can study the energy decay in a compact region
outside the black hole. As already mentioned, the spectrum of
stationary solutions is continuous, however we will show that
pseudo-stationary initial data constructed from a particular discrete
subset of stationary solutions, that of the resonant states, yields
long lasting configurations.

As described above, we construct initial configurations that
correspond to stationary solutions in the domain of dependence of the
region $r>2M+\epsilon$, with $\epsilon \lesssim 0.05M$.  To do so, we
integrate Eq.~\eqref{Eq:EigenvalueProblem} numerically in the region
$[R_{\rm in}=2M+\epsilon, R_{\rm out}]$ using a shooting algorithm
(see e.g.~\cite{Press86}).  As boundary condition we impose at
$r=R_{\rm out}$ the relation $u'=-|k| u$ with the constant $k$ defined
in Eq.~\eqref{eq.r.infty} to ensure that $u(r)$ falls exponentially at
large $r$ (see also~\cite{Megevand:2007uy} for more details). For the
left boundary condition we just impose $u(R_{\rm in})=1$, since the
equation is homogeneous so $u(r)$ can be rescaled afterwards. Finally,
$u(r)$ is simply set to $0$ for $r<R_{\rm in}$.   This introduces a
discontinuity in the scalar field configuration, but the discontinuous
jump is very small compared to the maximum absolute value of the
scalar field and does not seem to introduce any numerical artifacts.

Leaving the value of $\epsilon$ aside, these
initial configurations are characterized by the parameters $\ell$ and
$M\mu$, which parametrize the effective potential $M^2 V_{\rm
  eff}(r^*/M)$, together with $M\omega$ (see
Eqs. \eqref{Eq:TimeIndependentSchrodinger} and
\eqref{effectivepotential} above).  Here we will restrict ourselves to
the region $0<\omega^2<\mu^2$. The solutions with $\omega^2>\mu^2$ are
not localized close to the black hole and they will not be interesting
for our purposes here.  For the subspace with $0<\omega^2<\mu^2$ one
can distinguish the following cases:

\begin{enumerate}
 \item Configurations constructed from stationary solutions in the
   resonance band, i.e. with $V_{\rm eff}^{\rm
     min}<\omega^2<\rm{min}\left\{V_{\rm eff}^{\rm
     max},\mu^2\right\}$.
\begin{enumerate}
 \item Pseudo-stationary configurations constructed from {\em
   resonant} stationary solutions. We will call these {\em
   pseudo-resonant} states. \label{type.resonant}
 \item Pseudo-stationary configurations constructed from {\em
   non-resonant} stationary solutions. We will call these {\em
   non-resonant} states. \label{type.nonresonant.w>min}
\end{enumerate}
 \item Configurations constructed from stationary solutions outside
   the resonance band. \label{type.nonresonant.w<min}
\end{enumerate}

From now on we will also refer to these configurations as
type~\ref{type.resonant}, type~\ref{type.nonresonant.w>min} and
type~\ref{type.nonresonant.w<min}, respectively. Here we are assuming
that the effective potential does have a local minimum for some value
of $r^*$, i.e. Eq.~\eqref{condition_real} is fulfilled.  If that is
not the case, all the configurations will be of type
\ref{type.nonresonant.w<min}.

\begin{figure}[ht]
  \begin{center}
    \includegraphics[angle=270,width=0.8\textwidth,height=!,clip]{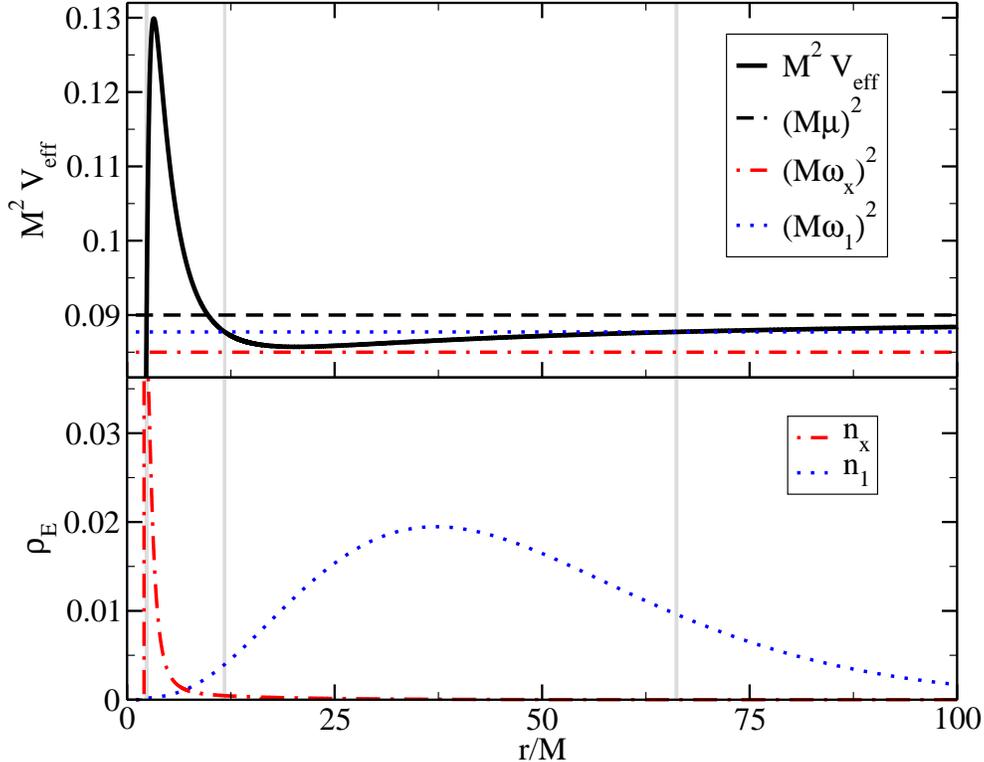}
    \caption{{\em Top panel:} Effective potential $V_{\rm eff}(r)$ for
      $\ell=1$ and $M\mu=0.3$; and $(M\omega_i)^2$ for each of the
      solutions shown in the bottom panel.  {\em Bottom panel:} Radial
      energy density $\rho_E(r)$ for the first ($n_1$) pseudo-resonant
      mode (type~\ref{type.resonant}) and a solution (labeled $n_x$) with
      eigenvalue (labeled $w_x^2$) below the local minimum of $V_{\rm eff}$
      (type~\ref{type.nonresonant.w<min}).  An arbitrary rescaling was
      chosen for each $\rho_E$ in this plot in order to improve
      visualization. The vertical lines intersect the roots of $V_{\rm
        eff}=w^2$, a total of three for the pseudo-resonant mode
      ($n_1$), and only one for the non-resonant mode (note that the
      root of the non-resonant mode is very close to the first root of
      the resonant mode, so the two corresponding vertical lines are
      indistinguishable in the figure).\label{f:Veff}}
  \end{center}
\end{figure}

Examples of type~\ref{type.resonant} and
type~\ref{type.nonresonant.w<min} configurations are shown in
Figure~\ref{f:Veff}, where we plot the effective potential, together
with two eigenvalues $(M\omega)^2$ (top panel), and the radial energy
density of the corresponding eigenfunctions (bottom
panel). The Figure is shown in
Eddington-Finkelstein coordinates, more appropriate for the numerical
evolution (see Section~\ref{sec:evolutions}).  For this particular
case we have chosen $\ell=1$ and $M\mu=0.3$, so that
Eq.~\eqref{condition_real} is fulfilled and a potential well is
present (that case was also considered in Section~\ref{sec:resonant}).
One would expect that such solutions will have an oscillatory behavior
in regions where $V_{\rm eff}<\omega^2$, with most of the scalar field
energy contained in those regions; and with a rapid decay outside. To
help guide the eye, we added vertical lines to the plots at the
locations where $V_{\rm eff}=\omega^2$ (classical turning points).  We
can see that most of the scalar field is located inside the
classically permitted regions ($V_{\rm eff}<\omega^2$).  In
particular, the type~\ref{type.nonresonant.w<min} configuration, for
which $\omega^2<V_{\rm eff}^{\rm min}$, is located mostly to the left
of the centrifugal barrier.  A type~\ref{type.nonresonant.w>min}
configuration has not been shown here, but we can look at these
configurations as somehow intermediate between
type~\ref{type.resonant} and type~\ref{type.nonresonant.w<min}.

We will now concentrate on type~\ref{type.resonant} configurations
which, as we will see in Section~\ref{sec:evolutions}, are the ones
that show long lasting scalar field distributions at a compact region
outside the black hole. We construct pseudo-resonant initial data
configurations for different values of $\ell$ and $M\mu$, and for each
pair of these parameters we study pseudo-resonant states up to the
fifth mode.

\begin{figure}[ht]
\begin{center}
\includegraphics[angle=270,width=0.6\textwidth,height=!,clip]{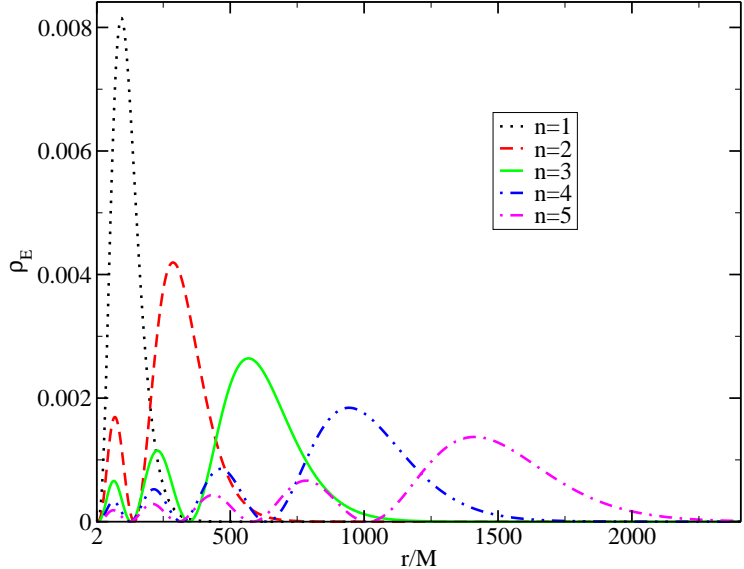}
\caption{Radial energy density $\rho_E$ at $t=0$ for different
  pseudo-resonant (type~\ref{type.resonant}) modes
  \mbox{$n=(1,2,3,4,5)$}, for the case $\ell=1$ and $M\mu=0.2$. Notice
  that the solution has been normalized so that we always have
  $E_{lm}=1$.\label{f:rho1}}
\end{center}
\end{figure}

\begin{figure}
\begin{center}
\includegraphics[angle=270,width=0.6\textwidth,height=!,clip]{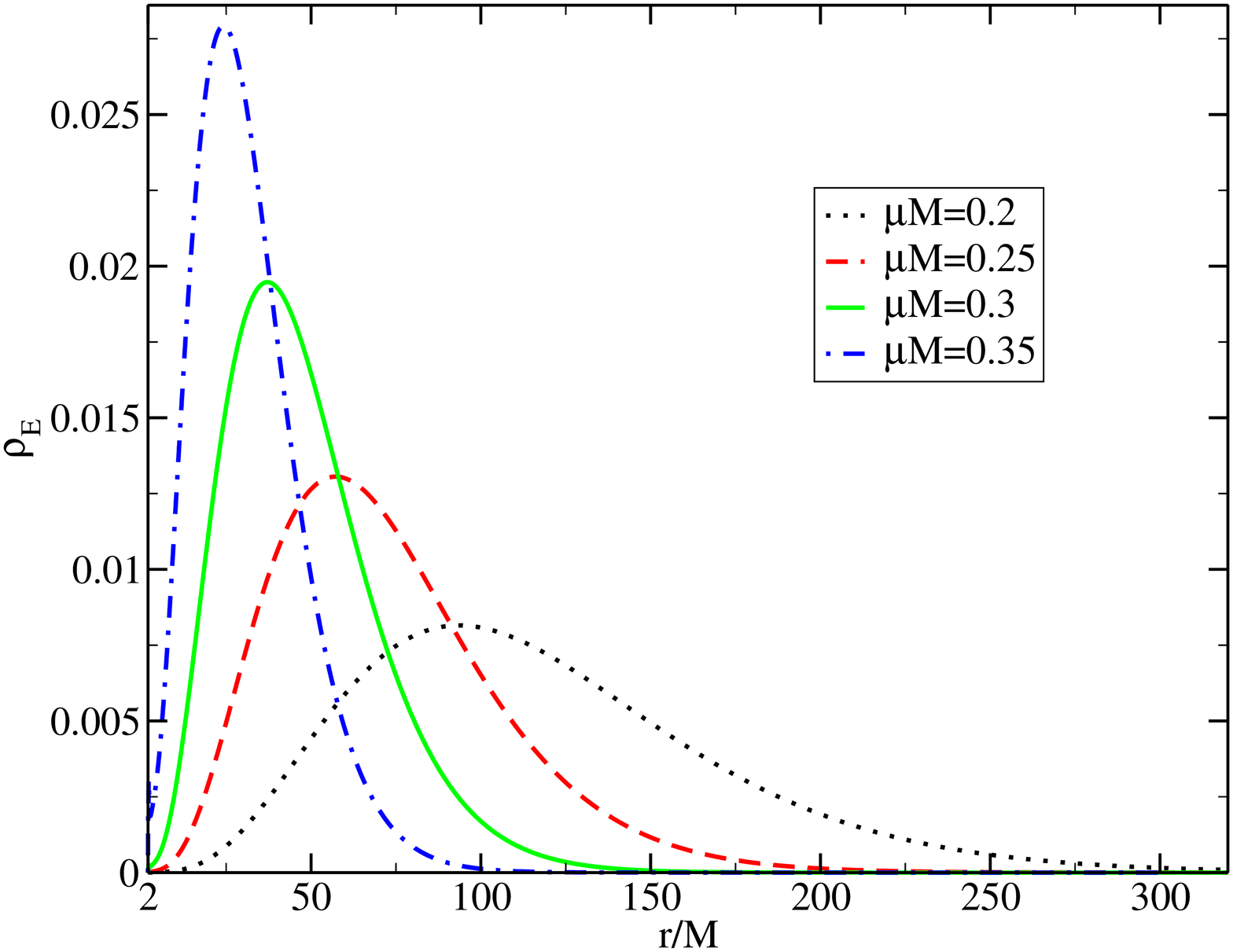}
\end{center}
\caption{Radial energy density $\rho_E$ for $\ell=1$ and the first
  pseudo-resonant (type~ \ref{type.resonant}) mode $n=1$, for
  configurations with different values of $M\mu$. The solution has
  been normalized so that $E_{lm}=1$. \label{f:rho2}}
\end{figure}

Table~\ref{table1} summarizes some properties of the pseudo-resonant
state initial data we have studied here. In what follows, $R_{99}$ is
defined as the radius of a sphere containing $99\%$ of the scalar
field energy $E$ at $t=0$, and gives a rough idea of the ``size'' of
each configuration. Note that $R_{99}$ increases as $M\mu$ decreases
and $n$ increases.  Some of these configurations are illustrated in
Figures~\ref{f:rho1} and~\ref{f:rho2}.  Figure~\ref{f:rho1} shows the
energy density distributions $\rho_E$ for the first five
pseudo-resonant modes $n=(1,2,3,4,5)$ for $\ell=1$ and $M\mu=0.2$,
while Figure~\ref{f:rho2} shows the first mode $n=1$ pseudo-resonant
configuration for different values of $M\mu$.

\begin{table}[ht]
\caption{For the case $\ell=1$, and different combination of
  parameters $M\mu$ and $n$ we show: (i) The ``width'' of the initial
  energy distribution $R_{99}$ and (ii) The eigenvalue $(M\omega)^2$.
  Notice that $R_{99}/M$ is given with three significant figure
  accuracy, while $(M\omega)^2$ is given with an accuracy up the last
  significant figure shown. All values shown correspond to
  pseudo-resonant (type~\ref{type.resonant}) states.}
\label{table1}
\begin{tabular}{l l c c c c }
\hline \hline
$n\;\;\;$ & $(M\mu)^2\rightarrow\;\;\;$ & $(0.2)^2=0.04\;\;\;$ & $(0.25)^2=0.0625\;\;\;$ & $(0.3)^2=0.09\;\;\;$ & $(0.35)^2=0.1225\;\;\;$ \\
\hline \hline
1  & $R_{99}/M$   & 279 & 173 & 116 & 79.7 \\
 & $(M\omega)^2$ & 0.03958216 & 0.061450 & 0.08773 & 0.1180\\
\hline 
2  & $R_{99}/M$   & 589 & 369 & 251 & 170 \\
 & $(M\omega)^2$ & 0.0398152 & 0.062037 & 0.089004 & 0.1205\\
\hline 
3  & $R_{99}/M$   & 1000 & 629 & 421 & 301 \\
 & $(M\omega)^2$ & 0.03989669 & 0.0622423 & 0.089449 & 0.1214\\
\hline
4  & $R_{99}/M$   & 1520 & 957 & 646 & 459 \\
 & $(M\omega)^2$ & 0.0399342 & 0.0623364 & 0.089652 & 0.12183\\
\hline
5  & $R_{99}/M$   & 2140 & 1350 & 917 & 651 \\
 & $(M\omega)^2$ & 0.0399545 & 0.0623871 & 0.089761 & 0.12204\\
\hline \hline
\end{tabular}
\end{table}

Note that the eigenvalues $w_n^2$ of the resonant modes accumulate at
$w^2=\mu^2$ from below. It seems likely that there is in fact an
infinite number of resonant modes $n$, although we have not
investigated in any detail if this is indeed the case.~\footnote{Note
  that in the case of bound states in quantum mechanics, effective
  potentials with an asymptotic behavior for $r\rightarrow\infty$ like
  the one in Eq.~\eqref{effectivepotential} usually have an infinite
  number of bound states.} The possible existence of modes with
arbitrary $n$ would be interesting, since the energy distribution
seems to spread further away from the black hole for larger $n$,
implying that arbitrarily wide resonant modes might be constructed.


\subsection{Numerical Evolution}
\label{sec:evolutions}

Having constructed a large set of initial data, we now study some
properties of their evolution.  The numerical evolution is performed
in penetrating coordinates, hence covering a region that reaches
inside the event horizon and avoiding the need to provide boundary
conditions there.  We use ingoing Eddington-Finkelstein coordinates
$(\bar{t},r,\theta,\varphi)$, with \mbox{$\bar{t}:= t -
  2M\ln(1-N(r))$}, and $N(r)$ defined in \eqref{Schwarzschild}. In
these coordinates the spacetime line-element takes the form
\begin{equation}
ds^2 = -\left(1-\frac{2M}{r}\right) \: d\bar{t}^2
+ \frac{4M}{r} \: d\bar{t} dr + \left(1+\frac{2M}{r}\right) \: dr^2
+ r^2 \: d\Omega^2 \; .
\end{equation}
This coordinate system covers a region that includes the black hole
interior $r\in(0,\infty)$, and is regular at the horizon.  Notice that
for the purposes of our simulations the Schwarzschild and
Eddington-Finkelstein times can be considered equivalent since at a
fixed radial coordinate $r$ one can easily see that $\Delta t = \Delta
\bar{t}$. Because of this, and in order to simplify the notation, from
now on we will drop the bar from the Eddington-Finkelstein time and
will refer to it simply as $t$.

In order to find numerical solutions to the Klein-Gordon equation, we
define a set of first order derivatives, and obtain a system of
evolution equations of the form \mbox{$\partial_t
  \vec{\Psi}+B\partial_r \vec{\Psi}=S$}, with $B$ a symmetric matrix
and $\vec{\Psi}$ the vector formed from first derivatives in space and
time of $\psi_{lm}$. See
references~\cite{Megevand:2007uy,Degollado:2009rw} for details.

The evolution equations are solved numerically using second order
finite differences in space, and evolving in time using a method of
lines with a third order Runge-Kutta integrator. Since we are using
horizon penetrating coordinates we set the left boundary $r_{\rm min}$
inside the event horizon, leaving the boundary conditions ``free'' at
that point (by using one-sided spatial differences), since that region
is causally disconnected from the exterior. On the other hand, the
right boundary $r_{\rm max}$ is set far away and all incoming modes
are set to zero there.  The code passed the standard convergence and
residual evaluation tests. Furthermore, two independent codes have
been used in this work, and comparisons give identical results up to
truncation error.

The resolution used varies depending on the initial data. Typically,
the region containing most of the scalar field (i.e. from
$r_{\rm min}$ to about $R_{99}$) is covered with at least 2,000, and
often as much as 32,000, grid points. Then, the domain is chosen so
that $r_{\rm max} \approx 4R_{99}$ or larger. Note that such high
resolutions are needed in some cases in order to calculate the
parameter $s$, defined below, with a relative error smaller than
one. This is mainly due to its proximity to zero.


\subsection{Results}
\label{sec:results}

We study the evolution of the scalar field $\psi_{lm}(t,r)$ for
different initial data configurations. Although our attention is
mainly focused on the evolution of the long lasting pseudo-resonant
states, we also consider other configurations for comparison. We start
by evaluating total energy loss and studying some spectral
characteristics of the different configurations by means of a time
Fourier analysis, and finish with more explicit considerations about
how long can such configurations last.


\subsubsection{Energy Decay and Spectral Analysis}
\label{sec:energyspectra}

In order to evaluate how long a scalar field configuration remains
confined in a compact region we evaluate the conserved energy
$E_{lm}$, defined in equation~(\ref{energy}), but we replace the upper
limit of integration with a finite value $R>R_{99}$, with typical
values of $R$ between $2R_{99}$ and $10R_{99}$.

\begin{figure}[ht]
  \begin{center}
    \includegraphics[angle=270,width=0.6\textwidth,height=!,clip]{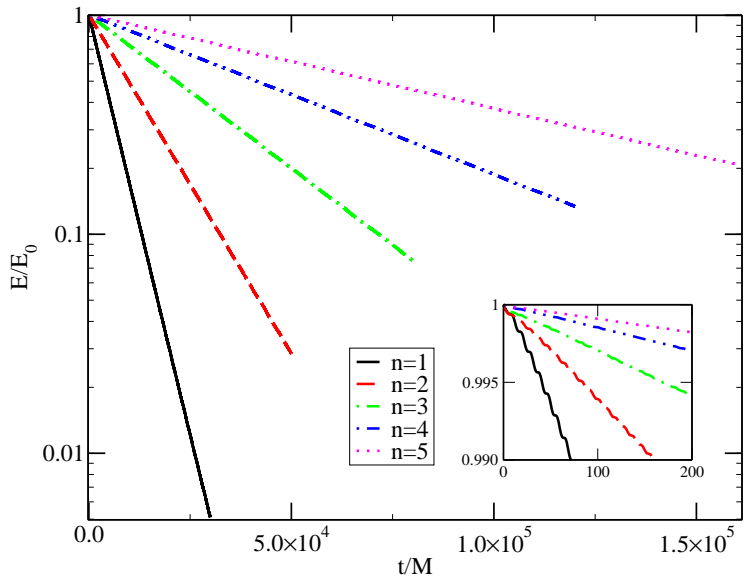}
  \end{center}
  \caption{Energy of the scalar field vs. time for the evolution of
    pseudo-resonant (type~\ref{type.resonant}) initial data with
    $\ell=1$, $M\mu = 0.35$ and modes $n=(1,2,3,4,5)$. As can be
    clearly seen some evolutions last longer than others, this is
    because we chose to run each case for a time length approximately
    proportional to the size of the scalar field distribution,
    $R_{99}$.}
\label{f:m_t0.2}
\end{figure}

\begin{figure}[ht]
  \begin{center}
    \includegraphics[angle=270,width=0.6\textwidth,height=!,clip]{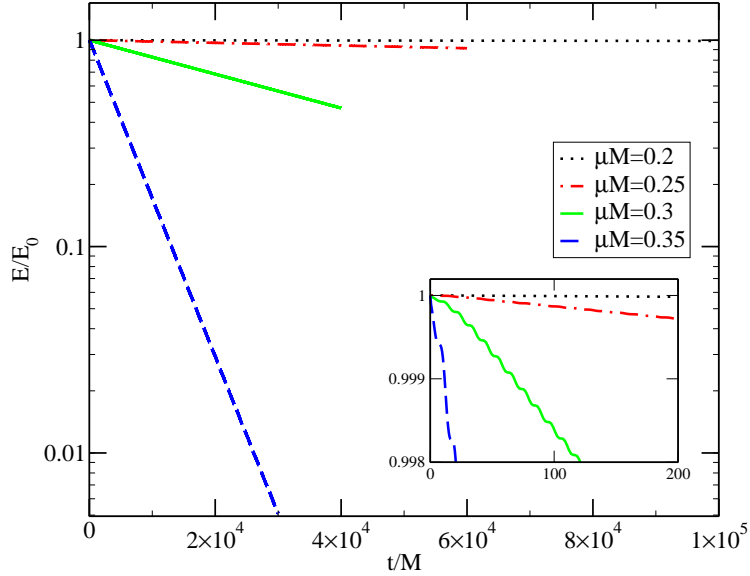}
  \end{center}
  \caption{Energy of the scalar field vs. time for the evolution of
    the initial data corresponding to the first ($n=1$)
    pseudo-resonant (type~\ref{type.resonant}) mode with $\ell=1$, and
    $M\mu=0.2$, 0.25, 0.3 and 0.35.}
\label{f:m_tw1}
\end{figure}

We begin by studying the evolution of the pseudo-resonant initial data
(type~\ref{type.resonant}) presented in table~\ref{table1}.
Figures~\ref{f:m_t0.2} and~\ref{f:m_tw1} show the evolution of the
integrated scalar field energy, $E(t)$ (we will drop the $\ell m$
sub-indexes from now on), for some of the parameters $M\mu$ and
$M\omega_n$ studied.  Notice that the energy is plotted on a
logarithmic scale, and shows a slow exponential decay of the form
\begin{equation}
E(t) = E_0 \exp(-s \: t/M) \; ,
\label{eq:decay}
\end{equation}
with $s$ constant, except for some very small oscillations that remain
during the whole evolution (these oscillations are noticeable in the
overlay frames in the figures). Given the exponential decay that
dominates the overall behavior, we can perform a linear fit of
$\ln(E/E_0)$ as a function of $t/M$ to calculate the parameter
$s$. These values are shown in Table~\ref{table} below. In all the
cases considered here no energy is seen to escape to infinity, but it
rather falls into the black hole. In Section~\ref{sec:times} below we
will evaluate explicitly the characteristic times for these
configurations. The results shown on the table correspond to the
evolution of pseudo-stationary initial data constructed from the
discrete spectrum of resonant stationary states
(type~\ref{type.resonant}).

\begin{table}[ht]
\caption{For the case with $\ell=1$, and each combination of
  parameters $M\mu$ and $n$, we show the slope $s$ of a linear fit of
  $\ln(E(t)/E_0)$, as described in the text.  The error in the slope
  is estimated either as one half the difference of the values
  obtained with the two higher resolutions used in each case, or as
  the error obtained from the linear regression, whichever is
  larger.} \label{table}
\begin{tabular}{ l  r|r  rr  rr  rr  }
\hline \hline 
$n$&&&& $(M\mu)^2$ &&&& \\
  && $(0.20)^2=0.04$\hfil\hfil && $(0.25)^2=0.0625$ \hfil\hfil&&
$(0.30)^2=0.09$\hfil\hfil && $(0.35)^2=0.1225$ \hfil \hfil\\
\hline 
1  && $(8.14\pm0.05)\t 10^{-8}$ && $(1.49\pm0.01)\t10^{-6}$ && $(1.857\pm0.002)\t10^{-5}$ && $(1.657\pm0.001)\t10^{-4}$ \\
2  && $(2.96\pm0.03)\t10^{-8}$ && $(5.53\t0.02)\t10^{-7}$ && $(7.20\pm0.05)\t10^{-6}$ && $(6.477\pm0.005)\t10^{-5}$ \\
3  && $(1.3\pm0.1)\t10^{-8}$ && $(2.479\pm0.005)\t10^{-7}$ &&$(3.24\pm0.02)\t10^{-6}$ && $(2.904\pm0.007)\t10^{-5}$ \\
4  && $(7\pm2)\t10^{-9}$ && $(1.30\pm0.01)\t10^{-7}$ && $(1.690\pm0.006)\t10^{-6}$ && $(1.506\pm0.002)\t10^{-5}$ \\
5  && $(5\pm4)\t10^{-9}$ && $(7.6\pm0.2)\t10^{-8}$ && $(9.83\pm0.01)\t10^{-7}$ && $(8.61\pm0.07)\t10^{-6}$ \\
\hline \hline
\end{tabular}
\end{table}

We have also evolved a couple of type~\ref{type.nonresonant.w>min}
configurations, that is, non-resonant pseudo-stationary states. Such
non-resonant states do not evolve in a quasi-stationary fashion,
instead they very quickly lose a significant portion of their energy,
some falls into the black hole while some is radiated toward
infinity. It is only after quite some time has passed that they reach
a quasi-stationary state, and with much less than their original
energy. Once they are in the quasi-stationary regime, the energy
falloff is very similar to that of the pseudo-resonant states.  Each
non-resonant state seems to evolve, at least after some time, as a
combination of pseudo-resonant modes.

To understand this behavior better we perform a spectral analysis.  We
calculate the discrete Fourier transform in time of the numerical
scalar field evolutions. More precisely, we calculate the magnitude of
the discrete Fourier transform in time of the field $\psi(t,r)$ at a
fixed spatial point (or points) $r=r_j$, located approximately at a
local extremum of $\psi(t=0,r)$, and pay attention to possible
differences derived from using different sample points $r_j$. The
explicit expression is
\begin{equation}
F[\psi(t)](f):= \left| A\; \sum\limits_{p}\; \psi(t_p,r_j)\exp(-2\pi i f t_p)\right|\;,
\end{equation}
where $A$ is a normalization constant and $t_p$ are the discrete time
values.

\begin{figure}[ht]
  \begin{center}
    \includegraphics[angle=270,width=0.6\textwidth,height=!,clip]{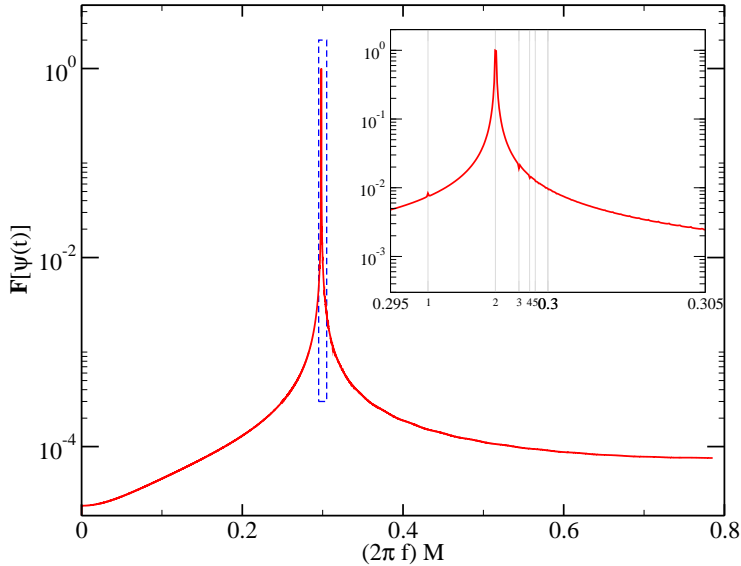}
  \end{center}
  \caption{Discrete Fourier transform in time vs. frequency for the
    evolution of type~\ref{type.resonant} data with $M\mu =0.3$,
    $\ell=1$, and frequency $w_2$ (corresponding to the second
    pseudo-resonant mode). The overlay figure shows in more detail the
    region contained in the dashed line rectangle. Numbers going from
    1 to 5 in the overlay figure denote the first five resonant
    frequencies. }
\label{f:fourier0}
\end{figure}

\begin{figure}[ht]
  \begin{center}
    \includegraphics[angle=270,width=0.6\textwidth,height=!,clip]{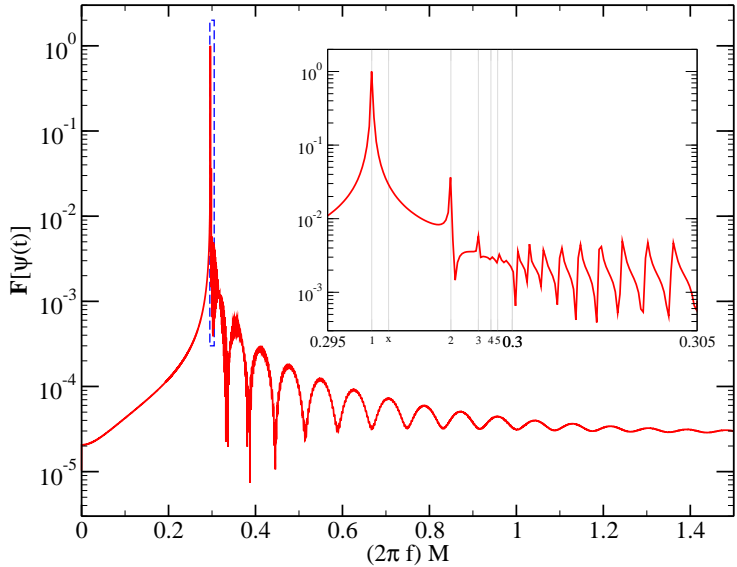}
  \end{center}
  \caption{Discrete Fourier transform in time for the evolution of
    type~\ref{type.nonresonant.w>min} data with $M\mu=0.3$, $\ell=1$,
    and frequency $w_x=0.29664794$. The overlay figure shows in more
    detail the region contained in the dashed line rectangle. Numbers
    going from 1 to 5 in the overlay figure denote the first five
    resonant frequencies, while the $x$ denotes $w_{\rm x}$.}
\label{f:fourier1}
\end{figure}

\begin{figure}[ht]
  \begin{center}
    \includegraphics[angle=270,width=0.6\textwidth,height=!,clip]{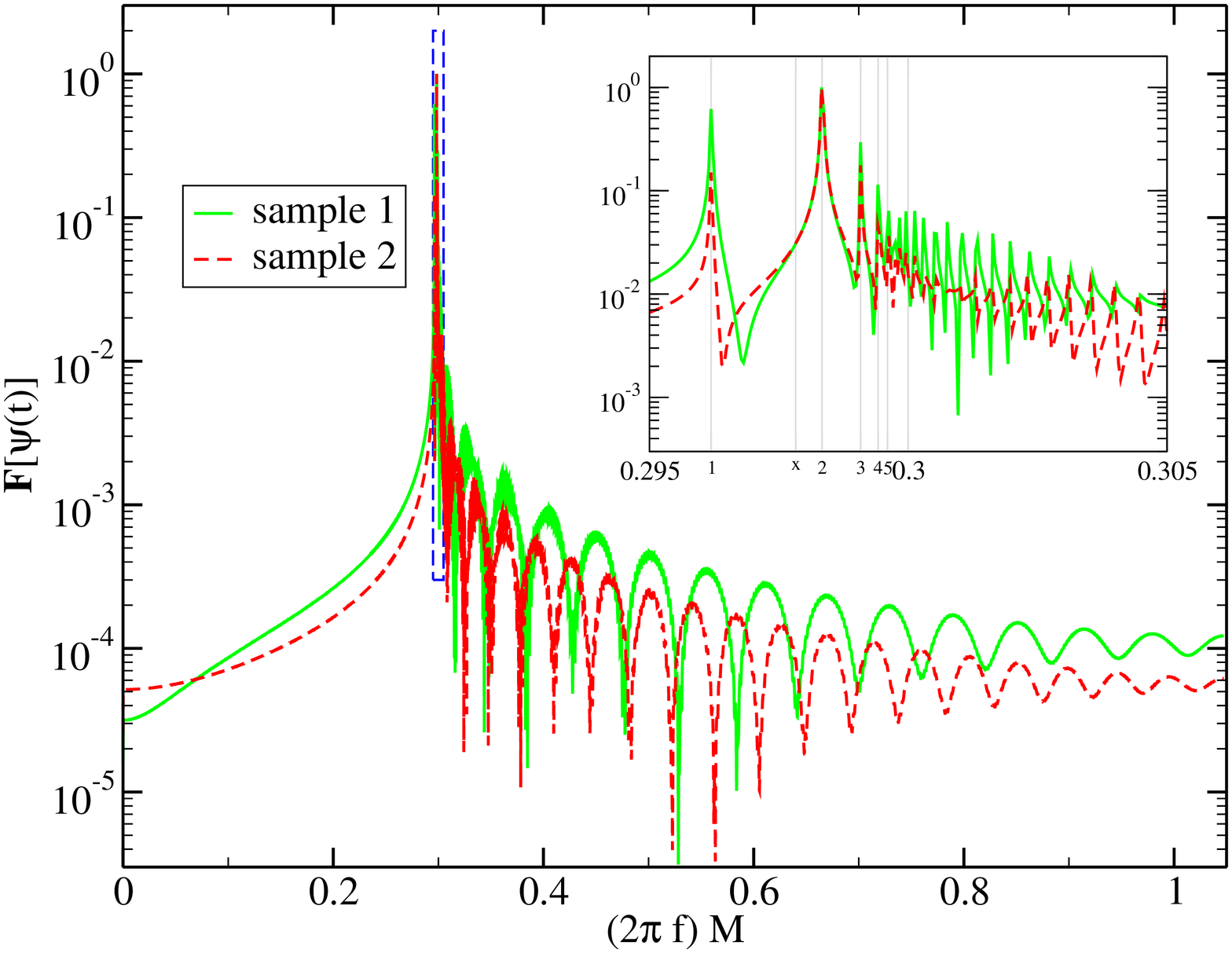}
  \end{center}
  \caption{Discrete Fourier transform in time for the evolution of
    type~\ref{type.nonresonant.w>min} data with $M\mu=0.3$, $\ell=1$,
    and frequency $w_x=0.29782545$. The overlay figure shows in more
    detail the region contained in the dashed line rectangle. Numbers
    going from 1 to 5 denote the first five resonant frequencies,
    respectively, while the $x$ denotes $w_{\rm x}$. The solid line
    curve (sample~1) corresponds to a sample point located close to
    the first extremum, while the dashed line curve (sample~2)
    correspond to one close to the second extremum.}
\label{f:fourier2}
\end{figure}

The Fourier transform of a (type~\ref{type.resonant}) pseudo-resonant
state is shown in Figure~\ref{f:fourier0}, and that of two
(type~\ref{type.nonresonant.w>min}) non-resonant ones is shown in
Figures~\ref{f:fourier1} and~\ref{f:fourier2}. For this example we
have chosen to show an $n=2$ pseudo-resonant state and two
non-resonant states with frequencies $\omega_x$ between the first and
second pseudo-resonant modes ($\omega_1<\omega_x<\omega_2$). These two
frequencies have been denoted with an $x$ in the overlay figures,
while numbers from 1 to 5 denote the first five resonant
frequencies. All cases correspond to $\ell=1$ and $M\mu=0.3$.

Figures~\ref{f:fourier0} and~\ref{f:fourier1} show the Fourier
transform at a single sample point $r_1$. While the results shown in
these two figures are almost independent of $r_1$, for the case
considered in Figure~\ref{f:fourier2} there were some small
differences depending on whether the sample point $r_j$ was chosen
close to the first or the second local extremum, so we show both
results in that case. Note however that the two curves in
Figure~\ref{f:fourier2} are qualitatively identical, and the peaks are
located at exactly the same frequencies, the conclusions we will make
from this figure are then independent of the sample point chosen.

Having mentioned a few technical details about the spectral analysis,
we now make some observations.  Consider first the spectrum of the
pseudo-resonant mode with frequency $\omega_2$ shown in
Figure~\ref{f:fourier0}. As might be expected for this configuration,
we see a single and well defined peak at exactly $\omega_2$. Note
however that this result is not so obvious a priori, since we are
actually evolving a truncated version of the actual stationary
state. Consider next the non-resonant configurations,
Figures~\ref{f:fourier1} and~\ref{f:fourier2}, with frequencies
$\omega_x$. We note that no peaks are present in the spectrum at the
frequency $\omega_x$, instead one can see clear peaks at different
resonant frequencies $\omega_n$.  Note also that, although not
strictly true for all cases, the peaks tend to be higher the closer
they are to $\omega_x$.  It would seem that, after some transient,
these non-resonant states evolve as a combination of pseudo-resonant
states.  In that sense, it seems that the pseudo-resonant states could
play a similar role to that played by the quasi-normal modes in the
study of perturbed black holes. This point should be investigated
further, but it is outside the scope of the present paper.

The Fourier transforms were calculated integrating over a
time interval that goes from $t=0$ to $t \approx 10^5M$. Such a large
interval is necessary to ensure that the final result has enough
frequency resolution to distinguish between the different resonant
frequencies $\omega_n$, which are very close to each other. This fact
also implies that we cannot integrate over shorter intervals in order
to evaluate possible time variations of the spectrum at the beginning
of the evolution, before reaching a quasi-stationary state.


\subsubsection{Characteristic Timescales}
\label{sec:times}

We now turn our attention back to the evolution of pseudo-resonant
(type~\ref{type.resonant}) configurations. Based on the results
outlined in table~\ref{table} we can be more explicit about how long
this states can last.  If the exponential regime observed is sustained
for all times, a characteristic time of these configurations, their
half-life time, will be given by $t_{1/2}:=\ln(2)M/s$, with $s$ the
parameter for the exponential decay of Eq.~\eqref{eq:decay}. Note that
in some of the cases considered we were able to evolve far beyond
$t_{1/2}$ thus corroborating, at least for those parameters, that the
exponential regime is indeed sustained (see
Fig. \ref{f:m_t0.2}).

Changing back from geometric units, and considering a black hole
mass of $10^8M_{\odot}$, we get a characteristic time of order
\begin{equation}
t_{1/2}\approx \frac{680}{s} \;\; \unit{seconds}\; .
\end{equation}
Replacing, for instance, the value $s=7\times10^{-9}$ that corresponds
to $\ell=1$, $M\mu=0.2$, $n=4$ (see Table~\ref{table}), we obtain
$t_{1/2} \approx 3,000$~years. This is still a very small time when
compared to cosmological time-scales, however, values of the parameter
$\mu$ motivated by dark matter scalar field models correspond to
$\hbar\mu \sim 10^{-24}\unit{eV}$ in physical units (see Introduction), or
equivalently $M\mu \sim 10^{-6}$ in geometric units, and noticing
how fast the parameter $s$ seems to decrease with decreasing $M\mu$
(see Table~\ref{table}), one might expect that configurations with
values of $M\mu$ motivated by dark matter scalar field models will
have half-life times of cosmological scales. We were, however, unable
to study cases with such small $\mu$, so the preceding argument is
purely hypothetical.

Note that extrapolating to values of $M\mu$ of order $10^{-6}$ from
the few values given in Table~\ref{table} which are of order
$10^{-1}$, would be extremely inaccurate.  If, however, we were to
proceed anyway with such an extrapolation, we would obtain $s$ of
about $6\times10^{-14}$, implying a value of $t_{1/2}$ of about $10^9$
years. Compare this value with the expected age of the universe,
$10^{10}$ years.~\footnote{In the parameter range studied, $s$
  decreases slightly faster than an exponential with decreasing
  $\mu$. For the extrapolation we assume an exponential that fits the
  points with smaller $M\mu$, $M\mu=0.2$ and $M\mu=0.35$.}

\begin{figure}
\begin{center}
\includegraphics[angle=0,width=0.8\textwidth,height=!,clip]{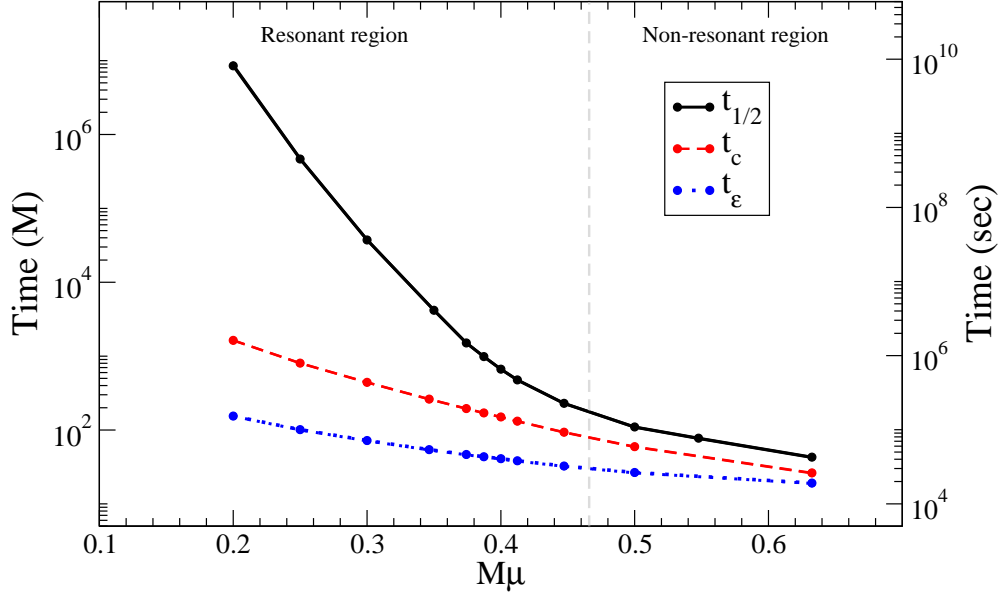}
\caption{Solid line: Half-life time of different scalar field
  configurations, $t_{1/2}:=\ln(2)M/s$.  Dashed line: ``Collapse
  time'', defined as the time in which a pressure-less gas sphere of
  size $R_{99}/2$ collapses in the gravitational field of the black
  hole, $t_c :=\sqrt{R_{99}^3/8GM}$.  Dotted line: Characteristic time
  defined at the beginning of Section \ref{sec:numerical},
  $t_\epsilon(2M+R_{99}/2)$. The left Y-axis shows these
  characteristic times in units of $M$, while the right Y-axis is
  expressed in seconds (assuming $M=10^8M_\odot$).
}\label{f:half_life}
  \end{center}
\end{figure}

Finally, for the purpose of comparison with the long lasting
(\ref{type.resonant}) pseudo-resonant configurations, we will consider
the evolution of configurations of type
(\ref{type.nonresonant.w<min}).  In the solid-line curve of
Fig.~\ref{f:half_life} we show the half-life time $t_{1/2}$ for both
pseudo-resonant states and states outside the resonance band for the
case $\ell=1$ and different values of $M\mu$.  A change in the slope
can be seen as soon as the critical mass $M\mu=0.466$ is reached,
showing that $t_{1/2}$ is indeed larger for type~(\ref{type.resonant})
states than for type~(\ref{type.nonresonant.w<min}) states.  Here
$t_{1/2}$ is expressed in the left vertical axis in units of $M$,
while on the right vertical axis it is expressed in seconds (assuming
$M=10^8 M_\odot$).

A quick inspection of Table~\ref{table1} shows that pseudo-resonant
states have a $R_{99}$ larger than that of non-resonant states,
i.e. they are wider.  One can argue that this is the reason why
pseudo-resonant states last longer.  In order to show that this is not
the case, we can make a comparison with some other characteristic
times.  For instance, we can compute the collapse time $t_c$, which is
the time in which a pressure-less gas configuration of size $\sigma$
collapses by its own gravity.  This is defined as
$t_c:=\sqrt{\sigma^3/ GM_g}$, where $M_g$ is the mass of the gas. In
our test field approximation, $M_g$ will be the mass of the black hole
$M_g=M$.  In order to compare $t_c$ with $t_{1/2}$ we found
appropriate to choose $\sigma$ as $R_{99}/2$.  This time is shown in
the dashed-line curve of Fig.~\ref{f:half_life}. As can be clearly
seen in this figure, $t_c$ is always smaller than $t_{1/2}$, and for
the particular case of pseudo-resonant states $t_{1/2}$ is orders of
magnitude larger than $t_c$.  Finally, for comparison, in the Figure
we also show (dotted-line curve) the characteristic time $t_\epsilon$
defined at the beginning of Section~\ref{sec:numerical}.


\section{Conclusions}
\label{sec:conclusions}

We investigated the existence of long-lasting scalar field
configurations surrounding a black hole.  A good motivation for this
study is the possibility that super-massive black holes at galactic
centers may represent a serious threat to the scalar field dark matter
models. This is because such black holes may in principle swallow an
hypothetical scalar field halo in cosmologically short times, which is
what in general will happen with arbitrary scalar field
distributions. However, we were able to find particularly long-lasting
scalar field configurations around a black hole.

As a first step, we have considered a relatively simple model and
looked for solutions to the Klein-Gordon equation on a Schwarzschild
space-time background. Although we do find stationary solutions for
the scalar field, we show that they are unphysical, in the sense that
their energy density integrates to infinity in a compact region just
outside the event horizon. However, we were able to find long-lasting,
quasi-stationary solutions of finite energy. This is done by evolving
initial data that was constructed by modifying slightly a particular
subset of the (unphysical) stationary solutions.  The solutions found
show as an overall behavior an exponential energy decay, caused by
scalar field leaking into the black hole, that in some cases can be
very slow.

The stationary solutions are obtained by solving a time-independent
Schr\"odinger-like equation with an effective potential. Hence, they
can be characterized by the properties of that potential. This fact is
strictly true for the stationary solutions, but interestingly we find
that the (physical) quasi-stationary solutions, for which the
Schr\"odinger-like equation no longer holds, are also characterized by
properties of that same effective potential. The cases of interest are
those in which the effective potential has a local minimum (a
potential well). The existence of this minimum depends solely on the
combination of parameters $M\mu$ and $\ell$. Although none of the
possible forms of the effective potential allow for bound states, the
existence of the mentioned well is enough to allow for the so-called
resonant states. These states are the ones that have proven useful for
constructing initial data that give rise to long-lasting
quasi-stationary configurations of finite energy.

It may be objected that in order to obtain the mentioned
quasi-stationary solutions one has to construct very particular
initial data, a situation that might be very unlikely in
nature. However, it seems that the crucial factor is only the
existence of the potential well. We showed that even when starting
with modified stationary solutions that are {\em not}\/ resonant,
after an initial abrupt energy loss, the late time behavior observed
is very similar to that of the resonant quasi-stationary solutions. In
fact, these solutions seem to evolve as a combination of the resonant
modes. Note that the value of $\mu$ for scalar field dark matter
models is expected to be given approximately by $\hbar
\mu=10^{-24}\unit{eV}$ in physical units, or $M \mu \sim 10^{-6}$ in
geometric units for a $10^8M_\odot$ black hole, which gives rise to
effective potentials with a local minimum for all values of $\ell$.

When evaluating the characteristic times of our solutions we find that the
longest lasting configurations last for times of the order of
thousands of years. Although this is cosmologically a very short time
we must note that, for technical reasons, we were only able to study
cases with relatively large values of $M\mu$, of order
$10^{-1}$. Noting how fast the characteristic times seem to increase
with decreasing $\mu$, it would not be unreasonable to expect that
configurations with $M\mu \sim 10^{-6}$ could last for cosmological
time-scales.  One could argue in the same lines about the ``size'' of
the scalar field distributions: The configurations studied extend only
up to a radius $R_{99} \approx 2,000M$, but again, this number increases
rapidly with decreasing $\mu$.

A few aspects of our study are still open for some improvement, and
will be addressed in future works. First, a self-gravitating scalar
field may be considered. Due to the extremely diluted nature of dark
matter halos, the test field (Cowling) approximation is very accurate
to some degree. However, given the halos large spatial extension, one
expects that important differences may appear globally as a cumulative
effect. Second, much smaller values of the parameter $\mu$, and much
larger scalar field distributions, will be needed to do a more
realistic representation of dark matter halos. The main difficulty in
dealing with such configurations is handling numerically the very
different scales, going from the regions close to the black hole to
galactic-size scales. Mesh refinement could be used to solve this
difficulty. Third, besides studying possible quasi-stationary or
long-lasting configurations with an already existing black hole, it
would be interesting to consider more dynamical scenarios such as the
formation and/or growth of the black hole and the possibility of
survival of the scalar field afterwards.  We plan to address these and
other issues in future works. However, the results presented here
already seem to indicate that it is indeed possible for scalar field
halos around super-massive black holes to survive for cosmological
time-scales.


\acknowledgments

We are grateful to the scalar field group at ICN-UNAM, for enlightening
discussions.  This work was supported in part by CONACyT through grants
82787 and 61173, and by DGAPA-UNAM through grant IN115311.  AD, JCD and MM
acknowledge DGAPA-UNAM for postdoctoral grants and AB and JB
acknowledge CONACyT for postdoctoral grants. OS was also supported by grant CIC 4.19 to Universidad Michoacana de San Nicol\'as de Hidalgo.


\bibliography{referencias}
\bibliographystyle{unsrt}


\end{document}